\documentclass[letterpaper]{article} 
\usepackage{aaai25}  
\usepackage{times}  
\usepackage{helvet}  
\usepackage{courier}  
\usepackage[hyphens]{url}  
\usepackage{graphicx} 
\usepackage{natbib}  
\usepackage{caption} 
\usepackage{algorithm}
\usepackage{algorithmic}
\usepackage{amsmath}
\usepackage{booktabs}
\usepackage{multirow}
\usepackage{multibib}
\newcites{sup}{Supplementary References}
\usepackage{newfloat}
\usepackage{listings}
\DeclareCaptionStyle{ruled}{labelfont=normalfont,labelsep=colon,strut=off} 
\floatstyle{ruled}
\newfloat{listing}{tb}{lst}{}
\floatname{listing}{Listing}
\title{Dr. Tongue: Sign-Oriented Multi-label Detection for Remote Tongue Diagnosis}
\author{
    Yiliang Chen\textsuperscript{\rm 1},
    Steven SC Ho\textsuperscript{\rm 1},
    Cheng Xu\textsuperscript{\rm 1},
    Yao Jie Xie\textsuperscript{\rm 1},
    Wing-Fai Yeung\textsuperscript{\rm 1},\\
    Shengfeng He\textsuperscript{\rm 2}, 
    Jing Qin\textsuperscript{\rm 1}
}
\affiliations{
    \textsuperscript{\rm 1}School of Nursing, The Hong Kong Polytechnic University, Hong Kong\\
    \textsuperscript{\rm 2}Singapore Management University, Singapore\\
    yiliang.chen@connect.polyu.hk
}

\usepackage{bibentry}

\begin{document}

\maketitle

\begin{abstract}
Tongue diagnosis is a vital tool in Western and Traditional Chinese Medicine, providing key insights into a patient's health by analyzing tongue attributes. The COVID-19 pandemic has heightened the need for accurate remote medical assessments, emphasizing the importance of precise tongue attribute recognition via telehealth. To address this, we propose a Sign-Oriented multi-label Attributes Detection framework. Our approach begins with an adaptive tongue feature extraction module that standardizes tongue images and mitigates environmental factors. This is followed by a Sign-oriented Network (SignNet) that identifies specific tongue attributes, emulating the diagnostic process of experienced practitioners and enabling comprehensive health evaluations. To validate our methodology, we developed an extensive tongue image dataset specifically designed for telemedicine. Unlike existing datasets, ours is tailored for remote diagnosis, with a comprehensive set of attribute labels. This dataset will be openly available, providing a valuable resource for research. Initial tests have shown improved accuracy in detecting various tongue attributes, highlighting our framework's potential as an essential tool for remote medical assessments.
\end{abstract}

\begin{links}
     \link{Resources}{https://github.com/tonguedx/tonguedx.}
\end{links}

\section{Introduction}

The global COVID-19 pandemic has accelerated the shift towards telemedicine, highlighting the critical need for reliable and efficient remote diagnosis methods~\cite{kichloo2020telemedicine,ohannessian2020global,portnoy2020telemedicine}. Within this context, tongue diagnosis, a cornerstone of Traditional Chinese Medicine, has gained renewed interest due to its potential to provide valuable insights into a patient's health through the analysis of tongue attributes~\cite{heo2022deep,hu2015variations,lee2016traditional,zhao2013differences,jiang2012integrating,lo2012study}. However, effectively incorporating tongue diagnosis into telehealth platforms requires the development of sophisticated automated recognition technologies capable of overcoming challenges such as inconsistent image quality and environmental conditions, to ensure precise and reliable diagnostics.

\begin{figure}[t]
  \centering
  \includegraphics[width=0.99\linewidth]{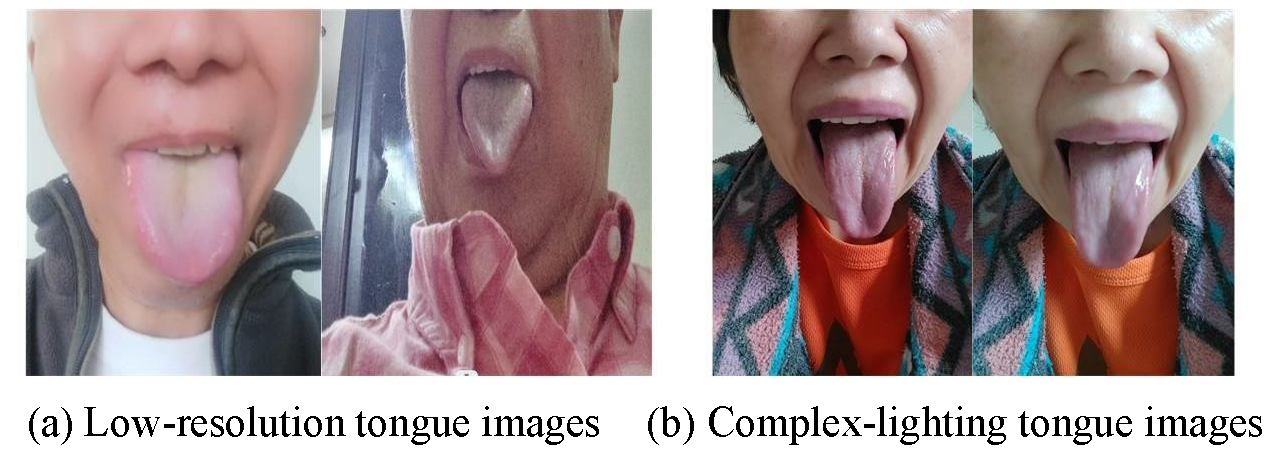}
  \caption{Challenges in tongue diagnosis imaging for telehealth: (a) low resolution and (b) complex lighting conditions.}
  \label{fig:1}
\end{figure}

The integration of tongue diagnosis into telehealth frameworks faces significant challenges, particularly in interpreting remotely captured images. Low-resolution cameras and beautification features on smartphones can compromise the quality of tongue images, which is critical for accurate diagnosis (Figure~\ref{fig:1}(a)). Additionally, variable lighting conditions can alter the perceived color and texture of the tongue, leading to potential misdiagnoses (Figure~\ref{fig:1}(b)). Images taken under different lighting conditions from the same individual can show marked differences in color and texture, complicating the diagnostic process.

To address these challenges, it is crucial to develop a comprehensive tongue image dataset. Despite numerous studies~\cite{li2018tooth,heo2022deep,hu2019automatic,wang2013statistical,wang2014research,jiang2022deep}, there is a lack of publicly available datasets that reflect the diverse conditions of telehealth settings. Existing datasets often rely on standardized equipment like the TFDA-1 tongue diagnosis device, which does not accurately represent telemedicine scenarios~\cite{shi2021clinical, jiang2021tongue,jiang2021application}. Moreover, many datasets lack detailed annotations and do not cover a wide range of tongue attributes necessary for comprehensive diagnosis~\cite{wang2014research, jiang2012integrating}.

To overcome these deficiencies, we propose to create a more comprehensive dataset that extends beyond mere tongue imagery. This dataset will include images captured in telemedicine scenarios, along with demographic details and comprehensive descriptions of tongue surface attributes, particularly tongue fur. Images will be collected using various cameras under different environmental conditions to ensure representative diversity. Our approach shifts the focus from identifying a single attribute to the concurrent recognition of multiple tongue surface attributes.

To address the challenges in tongue image analysis, such as environmental interference and posture variations, we propose a novel methodology comprising two main components: The Adaptive Tongue Feature Extraction module and the Sign-Oriented Attribute Detection network (SignNet). The Adaptive Tongue Feature Extraction module employs GroundingDINO~\cite{liu2023grounding} and SAM models~\cite{kirillov2023segment, ke2024segment} to precisely segment the tongue from complex backgrounds and align it to a standardized orientation. This standardization reduces the impact of environmental and postural variations, similar to object orientation methods~\cite{fu2008upright}, simplifying subsequent analysis.

Building on the standardized input, SignNet employs a multi-branch architecture to capture inter-attribute relationships effectively. The network consists of whole tongue, tongue body, and tongue edge sections, aligned with the attribute-sign relationships outlined in Table~\ref{tab:2}. To enhance inter-attribute learning, the design incorporates several key mechanisms.
The tongue body and edge sections integrate regional information, enabling the detection of location-specific features. The whole tongue branch extracts and integrates color and fur information, reorganizing color attributes into four categories and simplifying fur characteristics to presence or absence. This structured information is processed through an attention mechanism, which facilitates learning inter-relationships among various features across branches.

The attention mechanism further fuses information from different branches, enabling the network to recognize patterns such as the association between specific regions, colors, and fur characteristics. For example, it can detect that a pale tongue edge with no fur corresponds to particular diagnostic patterns.
This comprehensive design allows SignNet to capture the complex interrelationships between tongue attributes, considering their spatial distributions and mutual influences. Consequently, SignNet provides reliable support for remote tongue diagnosis, closely simulating the diagnostic practices of TCM practitioners. By learning these intricate relationships, our method achieves more accurate and comprehensive tongue attribute recognition, leading to more reliable and precise health assessments.

In summary, our contributions are threefold:
\begin{itemize}
  \item We create a diverse and comprehensive tongue image dataset tailored for telemedicine, including telehealth scenario images, demographic information, and multi-label annotations. This publicly accessible dataset addresses the common issue of restricted dataset availability in this field.
  \item We introduce a novel Sign-Oriented Attribute Detection framework that integrates adaptive tongue feature extraction with a Sign-oriented recognition Network (SignNet), enhancing the accuracy and robustness of automated tongue diagnosis across varied environmental conditions.
  \item We demonstrate the framework's adaptability to various imaging conditions, validating its efficacy and robustness in telehealth applications through experimental results.
\end{itemize}

\section{Related Work}
\subsubsection{Zero-shot Learning in Vision-Language Models.}

Zero-shot learning is a key feature in Vision-Language Models (VLMs), enabling recognition of new concepts without training~\cite{wang2022generalizing, xian2018zero}. CLIP~\cite{radford2021learning, yu2024beyond} pioneered this field through natural language supervision. Models like ALIGN~\cite{jia2021scaling} and Florence~\cite{yuan2023development} further enhanced performance and scalability, advancing tasks like object detection~\cite{bansal2018zero, zhu2019zero} and segmentation~\cite{zheng2021zero, bucher2019zero}. Unified architectures such as BLIP~\cite{li2022blip} and SimVLM~\cite{wang2021simvlm} show promise across vision-language tasks.

Recent advancements have pushed the boundaries of zero-shot capabilities in specialized domains. GroundingDINO~\cite{liu2023grounding} extended open-set object detection by using natural language prompts, while Segment Anything Models (SAM)~\cite{kirillov2023segment, ke2024segment} advanced zero-shot segmentation with a promptable interface that generates masks for arbitrary objects. These developments underscore the growing flexibility and generalization power of VLMs in novel scenarios. In our method, we leverage GroundingDINO and SAM to process tongue images within complex backgrounds, enhancing the robustness of our diagnostic framework.

\subsubsection{Tongue Images Dataset and Studies.}
The specialized domain of automated tongue diagnosis has received limited attention, largely due to its roots in traditional Eastern medicine rather than mainstream Western practices. A significant early contribution was made by Wang et al.~\cite{wang2013statistical}, who developed the first publicly accessible dataset for tongue diagnosis. However, this dataset was limited in scope, offering only basic demographic details like age and gender alongside the tongue images. Moreover, it has since become unavailable, creating a resource gap for researchers in this under-explored field.

Wang et al. also proposed several traditional machine learning techniques~\cite{pang2004computerized,zuo2004combination, zhi2007classification,huang2010tongue} to explore the relationship between abnormal tongue appearances and various diseases, focusing on aspects like color and shape analysis. As deep learning evolved~\cite{xu2022multi,jiang2024vrdone}, research in tongue image analysis began to address more complex attributes. Li et al.~\cite{li2018tooth} employed basic CNNs to recognize tooth-marked tongues, and another team~\cite{yuan2023development} used CNNs to investigate the connection between tongue images and stomach cancer, both as straightforward binary classification tasks.

More recently, Jiang et al.~\cite{jiang2022deep} advanced the field by using deep learning to identify multi-label tongue annotations. Their images, collected using the TFDA-1 tongue diagnosis instrument, achieved high accuracy due to the controlled imaging conditions, unaffected by angle or lighting variations. However, while this method is effective, it is not ideal for real-world telehealth scenarios where conditions are less controlled. To address this, we implemented a Sign-Oriented Multi-label Attributes Detection framework in our experiments, aiming to replicate the real diagnostic process and improve accuracy in diverse and realistic telehealth environments.

\begin{figure}[t]
  \centering
  \includegraphics[width=0.9\linewidth]{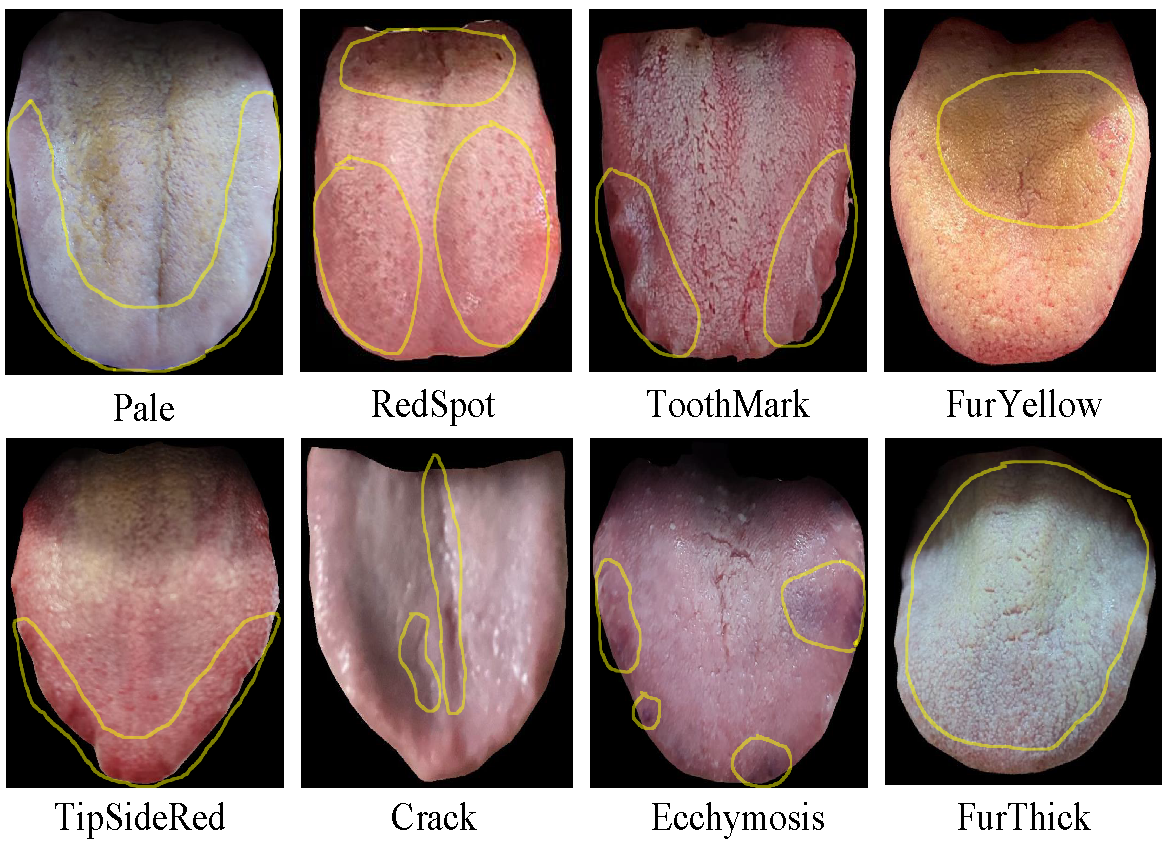}
  \caption{Cropped images with eight representative tongue attributes, with characteristics encircled in yellow.}
  \label{fig:2}
\end{figure}

\begin{figure}[tb]
  \centering
  \includegraphics[width=0.9\linewidth]{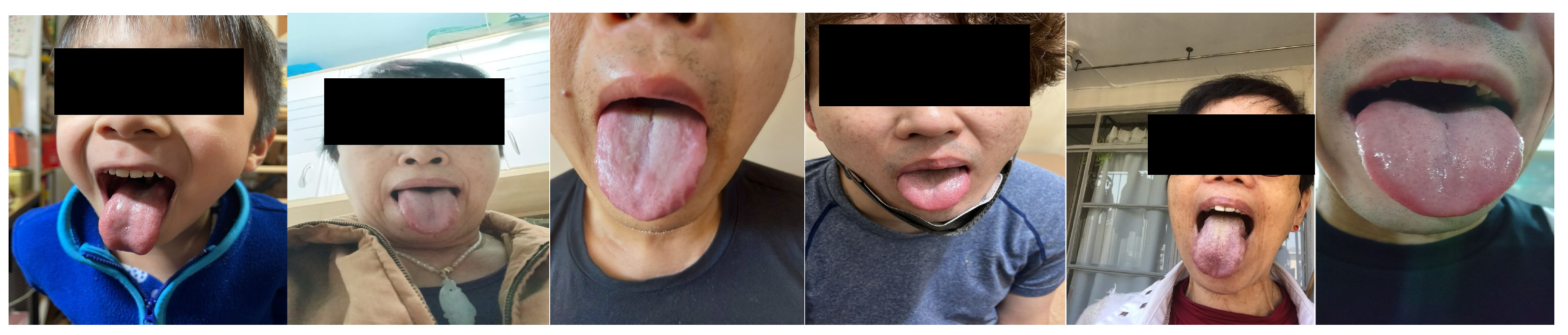}
  \caption{Original samples of tongue images from our dataset.}
  \label{fig:3}
\end{figure}

\begin{table}[t]
\centering
\begin{tabular}{lcccc}
\toprule[2pt]
Tongue attributes & No$_i$ & Yes$_i$ & No$_s$ & Yes$_s$  \\
\midrule
Pale & 4485 & 624 & 4083 & 567\\
TipSideRed & 2866 & 2243 & 2691 & 1959\\
RedSpot & 2678 & 2431 & 2457 & 2193\\
Ecchymosis & 4638 & 471 & 4219 & 431\\
Crack & 835 & 4274 & 754 & 3896\\
ToothMark & 2125 & 2984 & 1856 & 2794\\
FurThick & 136 & 4973 & 119 & 4531\\
FurYellow & 4276 & 833 & 3891 & 759\\
\bottomrule[2pt]
\end{tabular}
\caption{Tongue Attributes Statistical Distribution of the Proposed Dataset. No$_i$ denotes the number of images without the attribute, while Yes$_i$ denotes the number of images with the attribute. Similarly, No$_s$ and Yes$_s$ indicate the number of subjects without and with the attribute, respectively.}
\label{tab:1}
\end{table}

\section{The TongueDx Dataset}
Our tongue image dataset comprises a diverse compilation of images captured in natural settings using various smartphone and laptop cameras. This results in a dataset with inherent variability in environmental conditions, image resolutions, and tongue capture angles. Some cameras may have slight beauty filters enabled by default, while others do not. Additionally, the dataset includes images with different facial inclusions; some feature the entire face, while others only show a portion of the face.

For the annotation of tongue bounding boxes, we employed the GroundingDINO~\cite{liu2023grounding} model, which can accurately localize tongues in complex backgrounds based on the prompt text ``tongue". While this model provides precise annotations, it requires significant computational time for inference. To address its computational limitations, we used GroundingDINO's outputs as high-quality training data for a more lightweight network, significantly reducing detection time while maintaining accuracy.

We define classes for tongue images based on several representative tongue attributes, identified through consultations with five professional practitioners. We established eight fundamental tongue attributes: pale tongue color (Pale), reddish tip or edges (TipSideRed), red spots (RedSpot), ecchymosis (Ecchymosis), deep cracks (Crack), pronounced tooth marks (ToothMark), thick coating (FurThick), and yellow coating (FurYellow). Figure~\ref{fig:2} illustrates these attributes with yellow circles highlighting areas of interest. Each attribute is binary-classified, where $1$ indicates its presence. `Pale' suggests a lighter tongue body color; `TipSideRed' denotes tip redness; `RedSpot' refers to small, red dots in the coating; `Ecchymosis' indicates bruise-like discolorations; `Crack' points out significant fissures in the central area; `ToothMark' shows dental impressions along the edges; `FurThick' suggests a thick coating (normally transparent); and `FurYellow' denotes a yellowish coating. A tongue may exhibit multiple attributes simultaneously. These attributes were identified through in-person observations at a clinic, with corresponding images captured either at the clinic or participants' homes. Three professional practitioners completed the final annotation of surface attributes.

Our dataset comprises a total of 5109 images collected from 4650 unique subjects. Each subject have one or two images captured from different distances or angles. This variability is considered to capture a more comprehensive representation of the subjects within the dataset. For the subjects in our study, we present the number of occurrences of the attributes in Table~\ref{tab:1}. In addition to these attributes, we have also labeled other demographic data such as age and gender for each subject, laying the groundwork for future research. We showcase several samples of our dataset in Figure~\ref{fig:3}.


\section{Methodology}
In this section, we elaborate the architecture of our model comprehensively. We start by outlining the problem formulation of the multi-label tongue attribute recognition task. Following this, we introduce the proposed Sign-Oriented framework as shown in Figure~\ref{fig:5}. This method begins with an Adaptive Tongue Feature Extraction (ATFE) module to standardize tongue images and mitigate environmental factors, followed by a sign-oriented network (SignNet) that combines various signs to identify specific tongue attributes. This approach emulates the diagnostic process of practitioners, enabling thorough patient assessment.


\subsection{Formulation}
Let $(X,B,M,T)$ denote the tongue image dataset, where $X$, $B$, $M$, and $T$ represent images, bounding boxes, segmentation masks, and tongue attributes, respectively. Our approach consists of four steps: 1) Detection, where a lightweight network identifies the tongue's position within each image $x \in X$, predicting its bounding box $b \in B$; 2) Segmentation, using the detected bounding box as a prompt for SAM to generate a precise segmentation mask $m \in M$; 3) Upright Orientation, transforming the segmented tongue image to ensure an upright orientation; and 4) Attribute identification, identifying the basic attributes of the tongue. Formally, we aim to construct a multi-stage framework $f_\theta: x_n \to (b_n, m_n, r_n, t_n)$, where $\theta$, $x_n$, $b_n$, $m_n$, and $r_n$ represent model parameters, the input image, the predicted bounding box, the segmentation mask, and the upright orientation transformation, respectively. $t_n \in \{0,1\}^s$ is the binary label vector for attributes, with $s = |T|$ being the total number of possible attributes. Each image may exhibit multiple attributes simultaneously.

\begin{figure}[tb]
  \centering
  \includegraphics[width=0.9\linewidth]{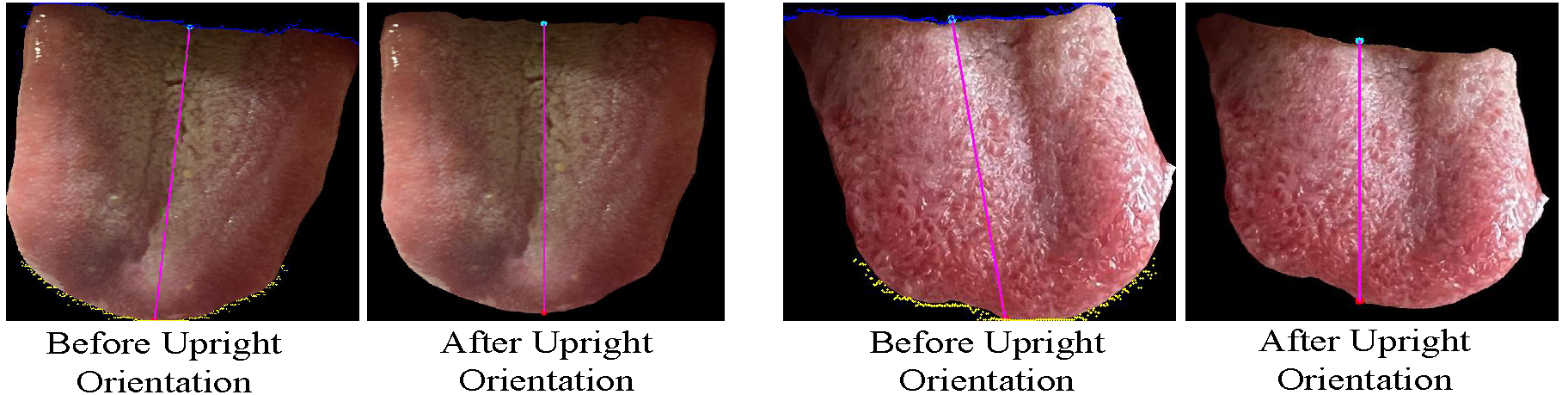}
  \caption{Two examples of the Tongue Image Upright Orientation Algorithm: Input image (Before Orientation) and the upright oriented tongue image (After Orientation).}
  \label{fig:4}
\end{figure}

\begin{figure*}[tb]
  \centering
\includegraphics[width=0.9\linewidth]{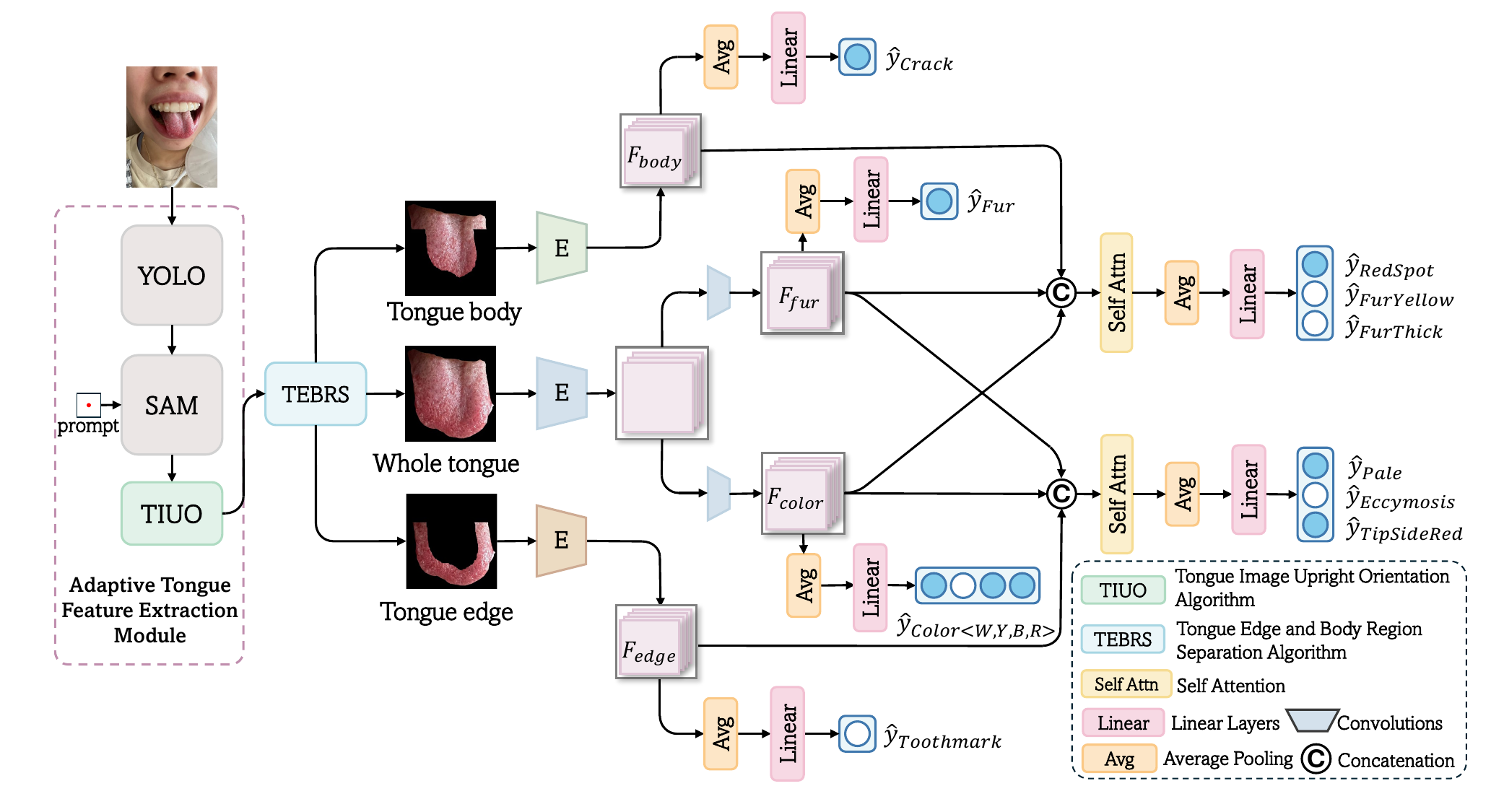}
\caption{Overview of our Sign-Oriented Attributes Detection Framework. The framework consists of three main stages: (1) Adaptive Tongue Feature Extraction (ATFE) module, which employs a detection network to locate the tongue, followed by Mobile-SAM (Segment Anything Model) for precise tongue segmentation. The segmented tongue image then undergoes upright orientation for standardization. (2) Tongue Edge and Body Region Separation (TEBRS), where the normalized tongue image is further separated into the tongue body and edge regions. (3) SignNet Pipeline, where the whole tongue, tongue body, and tongue edge images are fed into separate network branches to predict eight distinct tongue attributes.}
  \label{fig:5}
\end{figure*}

\subsection{Adaptive Tongue Feature Extraction Module}
\subsubsection{Tongue Detection and Segmentation.}
To accurately capture the tongue body in complex image backgrounds, we employ a multi-step approach. Initially, we utilize GroundingDINO to generate precise ground truth bounding boxes for each tongue image in our dataset. By inputting the text prompt ``tongue" and selecting the detection box with the highest confidence score, we obtained high-quality annotations. While GroundingDINO's accuracy proved sufficient, its computational demands as a large model made it impractical for real-time applications. To address this efficiency issue, we opted for YOLOv5-MobileNetV3~\cite{howard2019searching, jocher2022ultralytics} as our detection network to identify tongue positions. Given the simplicity of this single-object localization task, we can easily achieve over 99\% mAP performance. Subsequently, we input the predicted bounding box into MobileSAM~\cite{zhang2023faster} for segmentation, adding a click prompt at the box's center to enhance accuracy. Our observations showed minimal impact of different SAM variants on segmentation quality. Thus, considering the performance-efficiency trade-off, we choose MobileSAM for our implementation.

\begin{algorithm}
\caption{Tongue Image Upright Orientation Process}
\label{alg:1}
\textbf{Input:} \
Color images $X_1, X_2, ..., X_n$, \
Grayscale conversion function $G(\cdot)$,\
Contour smoothing functions $S_u(\cdot)$ and $S_l(\cdot)$, \
Tongue contour detection function $F(\cdot)$,\
Image rotation function $R(\cdot, \theta)$,\
Image translation function $T(\cdot, \Delta x, \Delta y)$, \
Margin $m$, smoothness $s$, and angle threshold $\alpha$
\begin{flushleft}
\textbf{Output:} \mbox{Uprighted tongue images $Y_1, Y_2, ..., Y_n$}
\end{flushleft}
\begin{algorithmic}[1]
\FOR{$i = 1$ to $n$}
\STATE $G_i = G(X_i)$
\STATE $U, L = F(G_i, \alpha)$ \hspace{2em} // Detect edges and filter points where angle between points $>$ $\alpha$
\STATE $U_s = S_u(U, G_i, s)$
\STATE $L_s = S_l(L, G_i, s)$
\STATE $T_{\texttt{tip}} = L_s[\lfloor |L_s|/2 \rfloor]$ \hspace{2em} // Find tongue tip point
\STATE $T_{\texttt{top}} = U_s[\lfloor |U_s|/2 \rfloor]$ \hspace{2em} // Find tongue top point
\STATE $\theta = \arctan2(T_{\texttt{tip}.y} - T_{\texttt{top}.y}, T_{\texttt{tip}.x} - T_{\texttt{top}.x})$ \hspace{2em} // Calculate rotation angle
\STATE $X_r = R(X_i, \theta)$ \hspace{2em}                  
                  // Rotate image
\STATE $\Delta x = \texttt{target}x - T{\texttt{tip}.x}$, $\Delta y = \texttt{target}y - T{\texttt{tip}.y}$
\STATE $X_t = T(X_r, \Delta x, \Delta y)$ \hspace{2em} // Translate image
\STATE $Y_i = \texttt{CropImage}(X_t, m)$
\ENDFOR
\STATE \textbf{return} $Y_1, Y_2, ..., Y_n$
\end{algorithmic}
\end{algorithm}

\subsubsection{Upright Orientation of Tongue.}

To standardize the orientation of tongue images for mitigating complex environmental factors and ensuring consistent feature extraction, we customize an approach to adjust the tongue to an upright position. This process ensures that the subsequent analysis is performed on uniformly oriented images, which is crucial to the accuracy and reliability of our tongue diagnosis system.

Our Tongue Image Upright Orientation Algorithm, as detailed in Algorithm~\ref{alg:1}, consists of several key steps. First, we convert the input color image to grayscale using the conversion operation $G(\cdot)$ for simplified processing. We then apply a contour detection approach $F(\cdot)$ to the grayscale image to identify the tongue's upper and lower contours, filtering out points where the angle of points exceeds a threshold $\alpha$. Note that our method assumes typical situations where the two contours must be located in the upper and lower regions of the image, and does not address extreme cases like completely upside-down or horizontal tongue images. This method efficiently traces the upper and lower contours of the tongue. To refine these contours for high-quality separation, we apply smoothing techniques using functions $S_u(\cdot)$ and $S_l(\cdot)$ for the traced contours. From these smoothed contours, we identify the tongue tip and top, which are the middle points of the lower and upper contours, respectively. The rotation angle is then calculated based on the line connecting the tongue tip and top, and the image is rotated by $R(\cdot, \theta)$. Afterwards, the image is translated using $T(\cdot, \Delta x, \Delta y)$ to align the tongue tip with a target position. Finally, the image is cropped with a specified margin $m$. In our experiments, we found that angle thresholds between $50^{\circ}$ and $70^{\circ}$ all produce similarly satisfactory results, with $60^{\circ}$ chosen for our final implementation. This process ensures that the tongue is consistently positioned in an upright orientation, facilitating more accurate and standardized feature extraction in subsequent steps, thus helping mitigate environmental variations such as lighting conditions and image qualities.

\subsection{Attribute-Sign Relationships Graph}

\begin{table}[tb]
\centering
\begin{tabular}{lcccc}
\toprule[2pt]
Surface attributes & Color & Shape & Region & Fur \\
\midrule
Pale & $\surd$ & & $\triangle$E & $\times$\\
TipSideRed & $\surd$ & & E & $\times$ \\
RedSpot & $\surd$ & & $\triangle$B & $\surd$ \\
Ecchymosis & $\surd$ & & $\triangle$E & $\times$ \\
Crack & & $\surd$ & B & \\
Toothmark & & $\surd$ & E & \\
FurYellow & $\surd$ & & $\triangle$B & $\surd$ \\
FurThick & $\surd$ &  & $\triangle$B & $\surd$ \\
\bottomrule[2pt]
\end{tabular}
\caption{Relationship between tongue attributes and their characteristic signs: color, shape, region, and fur.}
\label{tab:2}
\end{table}

After standardizing tongue images, we tackle the identification of eight basic surface attributes. However, using a single ResNet50 or ViT yields poor results for some attributes, particularly pale and ecchymosis in Table~\ref{tab:3}. This may stem from the network's lack of understanding of the specific areas to be identified. For instance, the pale attribute image in Figure~\ref{fig:2} only refers to the tongue region without fur. Although the fur area may appear pale in color, it is not included in the scope of pale attribute identification. This contradictory situation, where areas are mutually exclusive but similar in color, is likely to confuse the network.

To address this, we established a relationship table (Table~\ref{tab:2}) between the eight tongue attributes and four characteristic signs: color, shape, region, and fur. In the table, `$\surd$' indicates a definite connection, `$\times$' denotes mutual exclusivity, `$\triangle$' suggests a highly likely relationship and blank spaces indicate no connection.

Table~\ref{tab:2} shows that color is associated with six attributes, excluding crack and toothmark. Shape relates only to crack and toothmark. Region-wise, some attributes are area-specific: toothmark and tipsidered occur at the edge (E), crack in the body (B), while pale and ecchymosis likely appear at the edge ($\triangle$E). RedSpot, FurYellow, and FurThick likely appear in the body region ($\triangle$B). Regarding fur, redspot, furyellow, and furthick describe fur characteristics, while pale, ecchymosis, and tipsidered areas are definitely fur-free.

To effectively analyze these region-specific attributes, we segment the tongue into body and edge regions for more in-depth analysis. These segmented regions and the original image serve as inputs to our SignNet architecture. 


\subsection{Sign-Oriented Attributes Detection Network}
To realize the complex attribute relationships described in Table~\ref{tab:2} and simulate the real-world diagnostic process of Practitioners, we propose a multi-branch network architecture as shown in Figure 5. Our network consists of three main parts: the whole tongue analysis branch, the tongue body analysis branch, and the tongue edge analysis branch.
 
First, the whole tongue branch processes the entire tongue image, predicting fur color and the presence of fur. The color classification is achieved by combining multiple related attributes into four main categories: the White category is determined by the presence of Pale or Furthick attributes, the Yellow category corresponds to the FurYellow attribute, the Black category is associated with the Ecchymosis attribute, and the Red category is determined by the presence of TipSideRed or RedSpot attributes. This approach simplifies the learning process by transforming the difficult structural recognition problem into a simpler color recognition task while helping mitigate similar-appearing attributes under various imaging conditions. Also, fur is a fine-grained structure that is challenging to identify. By detecting the presence of fur, the network can initially capture the concept of fur and further learn based on the information in Table~\ref{tab:2} and other complementary data. The tongue body branch focuses on the central area of the tongue, with one of its branches dedicated to identifying the crack attribute that only appears in the body region. Similarly, the tongue edge branch analyzes the edge area of the tongue, with one branch identifying tooth marks and predicting the toothmark attribute specific to the edge. Subsequently, we employ attention mechanisms to comprehensively fuse the color and fur information from the whole tongue network with the tongue edge and tongue body network information, respectively. For feature fusion, we apply:
\begin{equation}
\begin{aligned}
F_{\text{cat\_i}} = [F_{\text{fur}}; F_{\text{color}}; F_{\text{i}}], \quad i \in \{\text{edge}, \text{body}\}
\end{aligned}
\end{equation}

\begin{equation}
Q_{\text{i}} = W_Q F_{\text{cat\_i}}, \quad K_{\text{i}} = W_K F_{\text{cat\_i}}, \quad V_{\text{i}} = W_V F_{\text{cat\_i}}
\end{equation}

\begin{equation}
A_{\text{i}} = \text{softmax}\left(\frac{Q_{\text{i}}K_{\text{i}}^T}{\sqrt{d_k}}\right)V_{\text{i}},
\end{equation}
where $F_{\text{fur}}$, $F_{\text{color}}$, $F_{\text{edge}}$, and $F_{\text{body}}$ represent features from fur, color, tongue edge, and tongue body, respectively. $F_{\text{cat\_i}}$ is the concatenated feature vector. $W_Q$, $W_K$, and $W_V$ are learnable weight matrices. The subscript $i$ denotes either the edge or body branch. Following the attention mechanism, we apply a Feed-Forward Network (FFN):
\begin{equation}
\text{FFN}(x) = \sigma(xW_1 + b_1)W_2 + b_2,
\end{equation}
where $W_1$, $W_2$, $b_1$, and $b_2$ are learnable parameters, and $\sigma(\cdot)$ represents the activation function. The complete feature fusion process can be expressed as:
\begin{equation}
\begin{aligned}
F'_{\text{i}} &= \text{LayerNorm}(F_{\text{i}} + A_{\text{i}}) \\
F''_{\text{i}} &= \text{LayerNorm}(F'_{\text{i}} + \text{FFN}(F'_{\text{i}})),
\end{aligned}
\end{equation}
where $i$ denotes either the edge or body branch. This process applies self-attention and a feed-forward network to each branch, with layer normalization after each step to enhance and stabilize the feature representations.

This network architecture, based on the attribute distribution in Table 2, simulates the TCM practitioners' diagnostic process. As our Attribute-Sign Relationships Graph illustrates, each tongue attribute has its corresponding regional, color, and fur characteristics. Our multi-branch structure reflects these relationships: edge-related attributes (pale, ecchymosis, tipsidered) and body-related attributes (redspot, furyellow, furthick) are processed in their respective branches. By fusing whole tongue color and fur information with edge and body features through attention mechanisms, our network effectively predicts basic tongue attributes for each region. This approach mimics the practitioners' diagnostic method, analyzing tongue attributes through multiple signs.


\begin{table*}[t]
\centering
\resizebox{\textwidth}{!}{%
\begin{tabular}{lcccccccccc}
\toprule[2pt]
\multirow{2}{*}{Model} & \multicolumn{2}{c}{Pale} & \multicolumn{2}{c}{TipSideRed} & \multicolumn{2}{c}{Spot} & \multicolumn{2}{c}{Ecchymosis} & \multicolumn{2}{c}{Crack} \\
\cmidrule(lr){2-3} \cmidrule(lr){4-5} \cmidrule(lr){6-7} \cmidrule(lr){8-9} \cmidrule(lr){10-11}
 & Acc & F1 & Acc & F1 & Acc & F1 & Acc & F1 & Acc & F1 \\
\midrule
Resnet50 & 88.31±0.56 & 32.68±2.04 & 72.94±1.17 & 66.19±0.81 & 78.06±1.31 & 76.44±1.26 & 90.75±0.58 & 24.72±2.83 & 84.92±0.86 & 91.17±0.55 \\
Vit & 89.45±0.13 & 1.61±2.20 & 66.64±0.56 & 43.41±1.57 & 60.92±1.34 & 50.90±5.15 & 90.84±0.00 & 0.00±0.00 & 83.46±0.00 & 90.99±0.00 \\
Resnet50+ATFE & 88.16±2.15 & 39.72±3.81 & 70.68±3.06 & 67.26±0.83 & 81.72±2.28 & 80.70±1.87 & 91.17±1.03 & 29.78±3.97 & 84.56±2.48 & 90.86±1.72 \\
TransFG & 89.61±0.00 & 0.00±0.00 & 64.02±1.11 & 31.08±7.12 & 62.68±3.30 & 43.63±8.56 & 90.84±0.00 & 0.00±0.00 & 83.46±0.00 & 90.99±0.00 \\
Laypersons & 48.94 & 18.54 & 49.39 & 45.62 & 51.06 & 49.42 & 49.72 & 14.77 & 47.71 & 59.59 \\
Practitioners & 86.82 & \textbf{43.81} & \textbf{76.20} & \textbf{71.64} & \textbf{82.91} & \textbf{81.81} & 85.14 & \textbf{38.14} & \textbf{87.71} & \textbf{92.46} \\
Ours & \textbf{90.21±1.10} & \textbf{46.00±5.80} & 74.08±1.72 & 68.22±1.06 & \textbf{82.35±0.70} & \textbf{81.09±0.84} & \textbf{92.13±0.48} & \textbf{38.02±3.05} & 86.23±0.88 & \textbf{91.90±0.56} \\
\midrule
\multirow{2}{*}{Model} & \multicolumn{2}{c}{ToothMark} & \multicolumn{2}{c}{FurThick} & \multicolumn{2}{c}{FurYellow} & \multicolumn{2}{c}{Average} & \multicolumn{2}{c}{\multirow{2}{*}{Jaccard}} \\
\cmidrule(lr){2-3} \cmidrule(lr){4-5} \cmidrule(lr){6-7} \cmidrule(lr){8-9}
 & Acc & F1 & Acc & F1 & Acc & F1 & Acc & F1 & & \\
\midrule
Resnet50 & 80.72±0.64 & 84.31±0.55 & \textbf{97.34±0.43} & \textbf{98.65±0.17} & \textbf{92.25±0.98} & 76.27±3.09 & 85.66±0.23 & 68.80±0.48 & \multicolumn{2}{c}{74.21±0.43} \\
Vit & 60.87±1.10 & 73.62±0.75 & 97.21±0.00 & 98.58±0.00 & 82.57±0.00 & 0.00±0.00 & 78.99±0.17 & 44.89±1.04 & \multicolumn{2}{c}{63.68±0.41} \\
Resnet50+ATFE & 80.12±2.50 & 83.63±1.73 & 97.34±0.09 & 98.65±0.05 & 92.11±0.44 & 70.92±0.80 & 85.73±1.12 & 70.92±0.80 & \multicolumn{2}{c}{74.86±0.99} \\
TransFG & 60.89±0.70 & 75.52±0.24 & 97.21±0.00 & 98.58±0.00 & 82.57±0.00 & 0.00±0.00 & 78.91±0.61 & 42.48±1.89 & \multicolumn{2}{c}{63.40±1.11} \\
Laypersons & 50.95 & 54.51 & 49.61 & 65.44 & 48.94 & 23.45 & 49.54 & 41.42 & \multicolumn{2}{c}{31.13} \\
Practitioners & \textbf{82.68} & \textbf{86.39} & 91.17 & 95.29 & 89.16 & 73.85 & 85.22 & \textbf{72.92} &  \multicolumn{2}{c}{\textbf{77.63}} \\
Ours & 79.24±0.98 & 82.82±0.61 & 97.14±0.19 & 98.54±0.10 & 92.07±0.35 & \textbf{76.41±1.00} & \textbf{86.68±0.30} & \textbf{72.87±0.63} & \multicolumn{2}{c}{75.83±0.37} \\
\bottomrule[2pt]
\multicolumn{11}{l}{\footnotesize Note: Acc columns show accuracy scores, and F1 columns show F1-scores. The Jaccard index provides a measure of overall performance across all attributes.}
\end{tabular}%
}
\caption{Quantitative comparison among different methods using five-fold cross-validation (values in mean\% $\pm$ std\%).}
\label{tab:3}
\end{table*}

\subsection{Loss Function}
We apply several loss functions to govern the training of SignNet in terms of three main tasks: classification of color, fur, and eight binary attributes. Each task is weighted to balance their contributions to the overall loss:
\label{eq:loss}
\begin{equation}
\mathcal{L}_{\text{total}} = w_{\text{color}} \mathcal{L}_{\text{color}} + w_{\text{fur}} \mathcal{L}_{\text{fur}} + \mathcal{L}_{\text{attr}}.
\end{equation}

Here, $\mathcal{L}_{\text{color}}$ uses sigmoid cross-entropy losses~\cite{xu2023pose} for the multi-label color classification, $\mathcal{L}_{\text{fur}}$ employs sigmoid cross-entropy loss for fur classification, and $\mathcal{L}_{\text{attr}}$ is a weighted sigmoid cross-entropy loss for the eight binary features:
\begin{equation}
\mathcal{L}_{\text{attr}} = \sum_{j=1}^{8} \alpha_j \mathcal{L}_{\text{BCE}}(y_j, \hat{y}_j),
\end{equation}
where $\alpha_j$ represents frequency-based weights for each binary feature, mitigating potential class imbalances. These weights are calculated as:
\begin{equation}
\alpha_j = \frac{median}{F_j}.
\end{equation}

Here, $m$ is the median frequency of all binary features in the training set, and $F_j$ is the frequency of binary feature $j$. 


\begin{table*}[t]
\centering
\resizebox{\textwidth}{!}{%
\begin{tabular}{lcccccccccc}
\toprule[2pt]
Model & Pale & TipSideRed & Spot & Ecchymosis & Crack & ToothMark & FurThick & FurYellow & Average & Jaccard  \\
\midrule
Resnet50 & 32.68$\pm$2.04 & 66.19$\pm$0.81 & 76.44$\pm$1.26 & 24.72$\pm$2.83   
         & 91.17$\pm$0.55 & \textbf{84.31$\pm$0.55} & 98.65$\pm$0.17 & 76.27$\pm$3.09 
         & 68.80$\pm$0.48 & 74.21$\pm$0.43\\
Resnet50+det     & 38.43$\pm$2.63 & 65.75$\pm$0.91 & 79.70$\pm$0.98 & 27.27$\pm$2.06 
         & 91.35$\pm$0.52 & 83.50$\pm$0.68 & 98.51$\pm$0.14 & 75.02$\pm$1.28 
         & 69.94$\pm$0.60 & 74.65$\pm$0.46 \\
Resnet50+seg     & 34.62$\pm$4.23 & 67.13$\pm$1.87 & 80.21$\pm$0.90 & 29.48$\pm$3.45 
         & 91.79$\pm$0.33 & 81.84$\pm$0.97 & 98.52$\pm$0.16 & 74.77$\pm$1.67 
         & 69.80$\pm$0.51 & 74.61$\pm$0.49 \\
Resnet50+upright & 39.72$\pm$3.81 & 67.26$\pm$0.83 & 80.70$\pm$1.87 & 29.78$\pm$3.97 
             & 90.86$\pm$1.72 & 83.63$\pm$1.73 & \textbf{98.65$\pm$0.05} & 76.77$\pm$1.55 
             & 70.92$\pm$0.80 & 74.86$\pm$0.99 \\
SignNet w/o color     & 40.65$\pm$2.04 & 66.10$\pm$0.94 & 81.05$\pm$0.37 & 34.94$\pm$4.47 
         & 91.10$\pm$0.38 & 82.23$\pm$0.86 & 98.55$\pm$0.22 & 75.99$\pm$2.53 
         & 71.33$\pm$0.38 & 74.62$\pm$0.60 \\
SignNet w/o fur     & 42.16$\pm$4.64 & 67.03$\pm$1.20 & 79.86$\pm$0.56 & 34.94$\pm$3.23 
         & 91.77$\pm$0.51 & 81.55$\pm$0.52 & 98.46$\pm$0.17 & \textbf{77.10$\pm$2.02} 
         & 71.61$\pm$0.64 & 74.74$\pm$0.80 \\
SignNet           & \textbf{46.00$\pm$5.80} & \textbf{68.22$\pm$1.06} & \textbf{81.09$\pm$0.84} & \textbf{38.02$\pm$3.05} 
               & \textbf{91.90$\pm$0.56} & 82.82$\pm$0.61 & 98.54$\pm$0.10 & 76.41$\pm$1.00 
               & \textbf{72.87$\pm$0.63} & \textbf{75.83$\pm$0.37} \\
\bottomrule[2pt]
\multicolumn{11}{l}{\footnotesize Note: All columns except the last one show F1-scores. The Jaccard index in the last column provides a measure of overall performance across all attributes.} \\
\end{tabular}%
}
\caption{Ablations of different model variants using five-fold cross-validation: F1-score and Jaccard (mean\% $\pm$ std\%).}
\label{tab:4}
\end{table*}

\section{Experiments}

\subsubsection{Compared Methods.}
To evaluate our method's effectiveness, we compared several models for tongue attribute recognition. Our baseline models include ResNet50 and Vision Transformer (ViT). We also reproduced the highest-accuracy method in Yuan et al.~\cite{yuan2023development}, which is based on TransFG~\cite{he2022transfg}. Other methods, such as Jiang et al.~\cite{jiang2022deep}, use bounding box annotations for subtle tongue features, making direct comparison impossible, and their code and datasets are also not publicly available. We further tested ResNet50+ATFE, where ATFE is our proposed tongue image standardization process (including detection, segmentation, and upright orientation), and our proposed SignNet.

Additionally, we conducted an experiment involving human evaluation. We invited three registered practitioners and five laypersons to classify eight simple attributes based on images from our test set, with their accuracy rates averaged. The laypersons received a brief training session on attribute recognition as illustrated in Figure~\ref{fig:2}. It is important to note that while our annotations were made through direct observation in a clinical setting, the human participants made their classifications solely from the images.

\subsubsection{Quantitative Results.}
As shown in Tables~\ref{tab:3}, the ViT-based models significantly underperform compared to ResNet50, with F1-scores for attributes like Pale and Ecchymosis approaching zero, indicating a fundamental lack of understanding of the task. This degradation can be attributed to ViT's limited capacity to capture high-frequency features. Similarly, the transformer-based TransFG architecture also exhibits comparable performance deficiencies.

Comparing ResNet50 with ResNet50+ATFE, we observe a significant improvement in F1-scores, with an average increase of about 2\%. Given the highly imbalanced nature of our dataset as shown in Table~\ref{tab:1}, we prioritize F1-score over accuracy. Our SignNet design further improves performance, achieving 1\% higher average accuracy and about 4\% higher average F1-score compared to the baseline ResNet50.

Additionally, all methods show particularly low F1-scores for pale and ecchymosis attributes. This could be attributed to the influence of overly bright or dark lighting conditions in the images. Notably, our network not only surpasses laypersons' accuracy but also approaches the performance of practitioners. This demonstrates the effectiveness of our proposed method in handling tongue attributes recognition.

\subsection{Ablation Study}
In this section, we present an ablation study on the efficacy of our main components in Table~\ref{tab:4}. ResNet50 serves as our baseline network. The Resnet50+det, Resnet50+seg, and Resnet50+upright models represent steps of our proposed ATFE module, corresponding to tongue detection, segmentation, and upright orientation, respectively. These components aim to standardize tongue image inputs. From Table~\ref{tab:4}, we can observe that images processed through the ATFE module show improvements in both average and F1-score metrics, demonstrating its effectiveness in mitigating varying imaging conditions.

Furthermore, we examine the impact of removing the color branch and fur branch from our proposed SignNet. The results labeled as `SignNet w/o color' and `SignNet w/o fur' illustrate the performance when these branches are individually removed. Notably, we can see that only when both branches are simultaneously included do we achieve the highest accuracy, showing their complementary roles in mitigating imaging variations. Therefore, color and fur branches play crucial roles in capturing tongue attributes and contributing to performance.

\section{Conclusion}
In this study, we developed a comprehensive tongue image dataset and proposed a novel framework for remote diagnosis, combining adaptive feature extraction and a sign-oriented recognition network. Our approach achieves near-practitioner performance despite dataset limitations. Future work aims to utilize all available labels, including demographics, to further enhance accuracy. 


\section{Ethical statement}
All procedures performed in this study with human participants were in accordance with ethical standards and approved by the Institutional Review Board of Hong Kong Polytechnic University. Informed consent was obtained from all individual participants. All tongue images were specially processed to protect personal privacy and prevent disclosure of identifying information. All data collected during the study was anonymized and securely stored, with access restricted to the research team.

\section{Acknowledgements} 
This work is partially supported by a Shenzhen-Hong Kong-Macao Science and Technology Plan Project (Category C Project) under the Shenzhen Municipal Science and Technology Innovation Commission (project no. SGDX20230821092359002), a grant for Collaborative Research with World-leading Research Groups of The Hong Kong Polytechnic University (project no. G-SACF), the Guangdong Natural Science Funds for Distinguished Young Scholars (Grant 2023B1515020097), the AI Singapore Programme under the National Research Foundation Singapore (Grant AISG3-GV-2023-011), and the Lee Kong Chian Fellowships.

\bibliography{aaai25}


\clearpage

\twocolumn[
    \begin{@twocolumnfalse}
    \begin{center}
    {\LARGE\bfseries Dr. Tongue: Sign-Oriented Multi-label Detection for Remote Tongue Diagnosis}

    \vspace{0.5cm}

    {\LARGE\bfseries —Supplementary Materials—}
    \end{center}

    \vspace{1cm}
    \end{@twocolumnfalse}
]

\section{Introduction}
In this document, we provide additional details and experiments to further examine our method for tongue image analysis. We begin by elaborating on our Tongue Body and Edge Region Separation Algorithm, which forms a crucial part of the input processing for our SignNet architecture as shown in Fig.~\ref{fig:5} in our main text, and is essential for analyzing region-specific attributes of the tongue. We then discuss the evaluation metrics used in our study, focusing on Accuracy, F1-score, and Jaccard index for our multi-label detection task, followed by a description of our five-fold cross-validation setting with a hold-out test set, explaining its importance for comparing our model with human assessments. Implementation details of our SignNet model are provided, including hardware specifications and optimization parameters. We also present a ROC curve analysis comparing different models' performance and discuss our model selection strategy. Additionally, we discuss the limitations of our study and outline future research directions. The document concludes with an ethics statement on responsible research.

\section{Tongue Body and Edge Region Separation}

To effectively analyze these region-specific attributes, we need to segment the tongue into body and edge regions. Algorithm~\ref{alg:2} details our method for this segmentation process. This algorithm operates on the upright-oriented tongue images obtained from our previous processing step (Algorithm~\ref{alg:1} in our main text).

We start with calculating an adaptive tongue edge width based on the image's diagonal and a predefined ratio parameter. This ratio (default: 0.191) allows us to flexibly adjust the tongue edge width for images with different aspect ratios, ensuring consistent separation across varied tongue shapes and sizes. Then, we apply function $F (\cdot)$ to detect the tongue region and create a full mask. Through morphological erosion using a kernel of size equal to the calculated tongue edge width, we create an eroded mask representing the inner region of the tongue. The edge mask is obtained by subtracting this eroded mask from the full mask, effectively isolating the peripheral region of the tongue.

To reduce noise and focus on the most relevant parts of the tongue edge, we remove the top fifth of the edge mask. The body mask is then derived by subtracting the processed edge mask from the full mask. This ensures that the body region excludes the peripheral areas captured in the edge mask. Finally, we apply these masks to the original image to extract the tongue body and edge regions. This is done through element-wise multiplication (denoted by $\odot$ in the algorithm) of the original image with the respective masks.

This adaptive separation approach, which accounts for image-specific dimensions, forms the basis for differentiating the regions mentioned in Table~\ref{tab:2} in our main text and is a crucial step in our SignNet architecture as shown in Figure~\ref{fig:5} in our main text. By separating the tongue into body and edge regions, our model is capable of focusing on the region-specific attributes, potentially improving the accuracy of our multi-label detection task.

\begin{algorithm}[t]
\caption{Tongue Body and Edge Region Separation}
\label{alg:2}
    \begin{flushleft}
\textbf{Input:}  \\
    Color images $X_1, X_2, ..., X_n$ \\
    Image height $h$ and width $w$ \\
    Tongue edge width ratio $r$ (default: 0.191) \\
    Tongue region detection function $F(\cdot)$ \\
\textbf{Output:} \\
    Tongue body regions $B_1, B_2, ..., B_n$ \\
    Tongue edge regions $E_1, E_2, ..., E_n$
    \end{flushleft}
\begin{algorithmic}[1]
\FOR{$i = 1$ to $n$}
    \STATE $e_w = \lfloor \sqrt{h^2 + w^2} \cdot r \rfloor$  //Calculate tongue edge width 
    \STATE $M_f = F(X_i)$  \hspace{3em} //Detect tongue region and create full mask 
    \STATE $M_e = \texttt{Erode}(M_f, \texttt{Ones}(e_w, e_w))$  \hspace{1em}//Eroded mask
    \STATE $M_{\texttt{edge}} = M_f - M_e$  \hspace{2em}//Get Tongue Edge mask
    \STATE $M_{\texttt{edge}}[0:\lfloor h/5 \rfloor, :] = 0$  //Remove top 1/5 edge area
    \STATE $M_{\texttt{body}} = M_f - M_{\texttt{edge}}$ \hspace{1em} //Get Tongue Body mask
    \STATE $E_i, B_i = X_i \odot M_{\texttt{edge}}, X_i \odot M_{\texttt{body}}$ 
\ENDFOR
\STATE \textbf{return} $B_1, B_2, ..., B_n, E_1, E_2, ..., E_n$
\end{algorithmic}
\end{algorithm}

\begin{table*}[!htbp]
\caption{Distribution of 8 tongue attributes across 5-fold cross-validation and hold-out test set in our TongueDx dataset. Values represent the count of occurrences for each attribute in each fold and in the test set.}
\centering
\label{tab:5}
\begin{tabular}{lcccccccc}
\toprule[2pt]
Fold & Pale & TipsideRed & Spot & Ecchymosis & Crack & Toothmark & Furrthick & Furellow \\
\midrule
Fold1 & 104 & 359 & 407 & 80 & 682 & 493 & 823 & 148 \\
\midrule
Fold2 & 102 & 399 & 402 & 81 & 723 & 495 & 814 & 137 \\
\midrule
Fold3 & 120 & 376 & 376 & 77 & 701 & 442 & 816 & 124 \\
\midrule
Fold4 & 114 & 354 & 403 & 81 & 717 & 475 & 825 & 143 \\
\midrule
Fold5 & 91 & 386 & 421 & 76 & 704 & 488 & 825 & 128 \\
\midrule
TestSet & 93 & 369 & 422 & 82 & 747 & 541 & 870 & 156 \\
\bottomrule[2pt]
\multicolumn{9}{l}{} \\
\end{tabular}
\end{table*}

\section{Evaluation Metrics}
Our task of identifying 8 tongue attributes is basically a multi-label detection problem. In the localization network, almost any type of network can easily achieve 99\% mAP as it is a simple single-object localization task. Therefore, we shift our main focus to the evaluation on the multi-label recognition aspect. We choose three representative metrics for evaluation: Accuracy, F1-score, and Jaccard index. Their formulas are as follows:

\begin{itemize}
\item Accuracy:
\begin{equation}
Accuracy = \frac{TP + TN}{TP + TN + FP + FN},
\end{equation}
where TP, TN, FP, and FN denote the True Positive, True Negative, False Positive, and False Negative, respectively.

\item F1-score:
\begin{equation}
F1 = 2 \cdot \frac{Precision \cdot Recall}{Precision + Recall},
\end{equation}
where Precision is the ratio of correctly predicted positive observations to the total predicted positive observations, and Recall is the ratio of correctly predicted positive observations to all observations in the actual class.

\item Jaccard index:
\begin{equation}
J = \frac{|A \cap B|}{|A \cup B|},
\end{equation}
where A and B are the predicted and ground truth label sets, respectively.
\end{itemize}

In addition to these metrics, we also report the average accuracy and average F1-score across all 8 attributes.

\section{Five-fold Cross-validation Setting}
In our experiment, we employ a five-fold cross-validation approach with a hold-out test set to evaluate our model on the TongueDx dataset. This approach involves setting aside a fixed portion of the data as a separate test set, while the remaining data is divided into five folds for training and validation. We chose this method over the alternative approach (where the entire dataset is divided into five equal folds, and the test set rotates through the entire dataset) because we need to compare our model with assessments by professional practitioners and laypersons. This comparison is necessary as a majority of tongue diagnosis studies utilize either simple models, have undisclosed architectures, or employ different labeling schemes, making direct comparisons challenging. However, the alternative approach is impractical for human evaluation as it would require too much time and effort from human assessors. Our chosen strategy ensures reliable comparison between our model and human performance while maintaining the benefits of cross-validation in the training process.

Each of our five folds consists of 751 subjects, with either 842 or 843 images. It's important to note that there is no subject overlap between the folds, ensuring the integrity of our evaluation process. As for the separate test set, it comprises 895 unique subjects, with each subject represented by exactly one image. This single-image-per-subject approach in the test set was specifically designed to facilitate fair comparisons with assessments by physicians or laypersons. By limiting each subject to one image, we mitigate the risk of recognition based on non-tongue-related features, which could potentially bias the accuracy of our results. Furthermore, we ensure that the subjects and images in the test set do not overlap with those in the training and validation sets.

To provide a comprehensive overview of our dataset's composition, we analyze the distribution of the eight key tongue attributes across the five folds. Table \ref{tab:5} presents a detailed breakdown of the occurrence count for each tongue attribute across the five folds. Each fold contains either 842 or 843 tongue images, while the separate test set comprises 895 tongue images. The distribution of tongue attributes remains relatively consistent across all five folds, with some minor variations. This consistency is vital for ensuring that each fold provides a representative sample of the overall dataset, thereby contributing to the robustness of our cross-validation process.

\section{Implement Details}
Experiments were conducted using an NVIDIA GeForce RTX 3090 Ti GPU. We employed the AdamW optimizer with an initial learning rate of 0.0002, which was linearly decayed during training. The training process utilized a batch size of 32. All input images, for both training and testing phases, were resized to a uniform dimension of 256 × 256 pixels. In Equation 6 of our loss function in our main text, we empirically set $w_{color}$ and $w_{fur}$ to 1 and 0.6, respectively.

\begin{figure}[tb]
  \centering
  \includegraphics[width=0.9\linewidth]{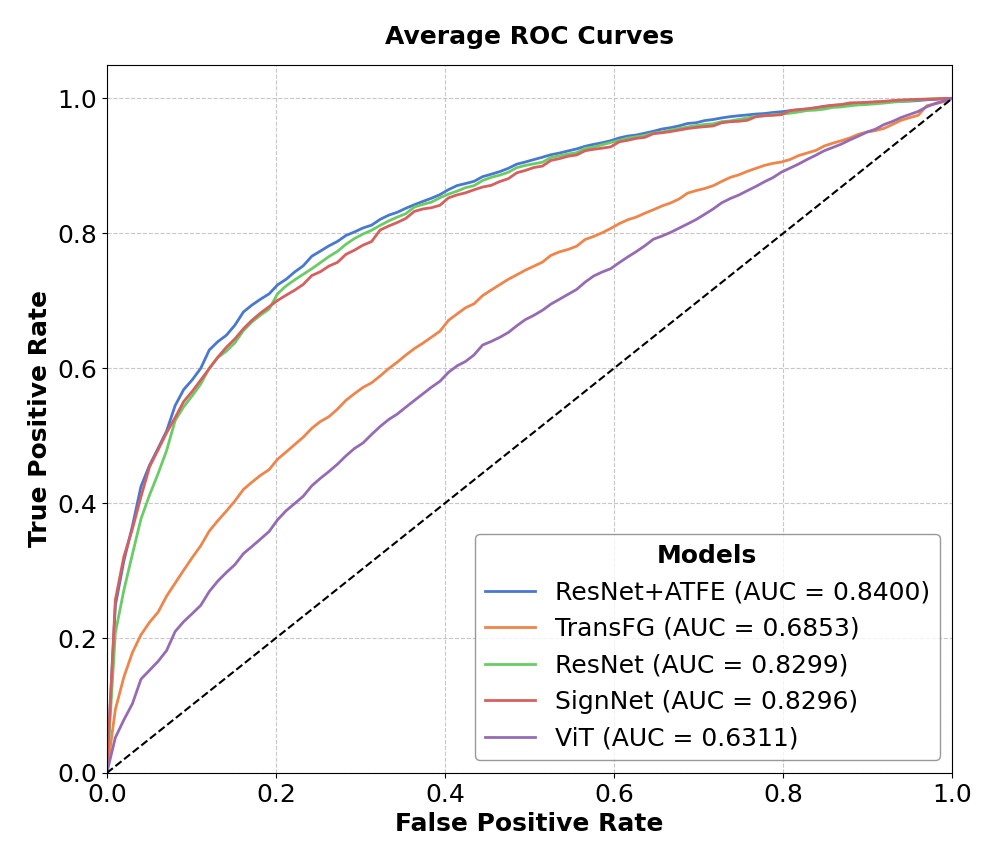}
  \vspace{-3mm}\caption{Receiver Operating Characteristic (ROC) curves comparing the performance of different models for tongue attribute recognition. The plot demonstrates that our proposed method shows comparable performance to other models across various tongue attributes.}\vspace{-3mm}
  \label{fig:6}
\end{figure}

\section{ROC Curve Analysis}
For a more comprehensive comparison, we also plotted ROC curves of different methods. As depicted in Figure~\ref{fig:6}, our proposed methods, ResNet50+ATFE (Adaptive Tongue Feature Extraction module) and SignNet, show performance closely aligned with the baseline ResNet across various tongue attributes, while outperforming other models. The performance of our proposed methods is fairly comparable to the baseline ResNet. This close alignment in performance can be attributed to our model selection strategy and evaluation methodology. 

It is worth noting that our model selection criteria prioritized the F1 score. Specifically, we selected the model that achieved the highest average F1 score on the validation set. This choice reflects our emphasis on balancing precision and recall, which is crucial for the tongue attribute recognition task. We consider all eight tongue attributes equally important for the overall task, hence the use of F1 score as the primary evaluation metric better reflects the model's performance in practical applications. Moreover, to maintain consistency and fair comparison, we used a uniform classification threshold of 0.5 for all attributes, rather than optimizing thresholds individually for each attribute to maximize F1 scores.

Nevertheless, our method still achieves sufficiently good performance in terms of accuracy in Table~\ref{tab:3} in our main text and AUC in Figure~\ref{fig:6}. This indicates that even without optimizing for these metrics specifically, our models maintain competitiveness across multiple evaluation dimensions. The ROC curves clearly demonstrate that our proposed methods offer a balanced trade-off between sensitivity and specificity across different tongue attributes.

These results highlight the robustness and versatility of our approach. While there is room for potential optimization through threshold tuning, the current results already prove the effectiveness of our method in handling the complex task of tongue attribute recognition. Future work could explore how to optimize performance for specific attributes or application scenarios while maintaining the overall balance of the model.

\section{Limitations and Future Work}

\subsection{Stability Issues in Results}
As demonstrated in Tables~\ref{tab:3} and~\ref{tab:4}, our approach exhibits considerable variability in performance, particularly evident in the high variance of five-fold cross-validation results for Pale and Ecchymosis attributes. This instability can be primarily attributed to the inherent data imbalance in our dataset. Specifically, as illustrated in Table~\ref{tab:5}, the instances of Pale and Ecchymosis attributes are significantly underrepresented, resulting in substantial fluctuations in performance metrics where even a single misclassification can dramatically impact the overall results. To address this limitation, our future work will focus on expanding the dataset with additional samples of these underrepresented attributes, aiming to achieve a more balanced distribution and, consequently, more robust and stable performance metrics.

\subsection{Potential of Color Correction Methods}
While our study advances tongue image analysis, it's important to acknowledge certain limitations. Our method may face challenges under extreme lighting conditions, potentially leading to reduced performance, especially for attributes sensitive to illumination changes like 'Pale' and 'Ecchymosis', as also evident in our demo.mp4 demonstration. These extreme lighting conditions can severely distort or obscure the visual attributes essential for accurate diagnosis, effectively compromising the integrity of the attribute information in the images. To address these illumination-dependent variations, incorporating color correction techniques could be a promising direction for future work. Specifically, methods like those proposed in \citesup{bianco2019quasi, Afifi_2019_CVPR, Afifi_2021_CVPR, Afifi_2020_CVPR} could help normalize tongue images across different lighting conditions.

Actually, this challenge has been previously explored by a prestigious tongue research team~\citesup{5570961}. They use a Munsell Colorchecker with multiple color patches as a reference target, capture the colorchecker under the same imaging conditions as tongue images, and measure its device-independent sRGB values using a spectrophotometer. By comparing the actual colors of the colorchecker with the captured colors, they derive a transformation matrix to correct all tongue images.

However, implementing such methods in our approach presents significant challenges, primarily because our dataset consists of retrospective data, making it impractical to apply the aforementioned color correction techniques. Additionally, real-world scenarios present more diverse and challenging conditions, including more extreme lighting variations. Therefore, conventional color correction methods designed for natural images may not be directly applicable in this context. The challenge is further complicated by the fact that the tongue's surface presents unique complexities due to its non-Lambertian properties, intricate 3D structure, and varied textures, with different regions reflecting light in distinct ways. Evaluating the effectiveness of color correction on such a complex surface would also require further experimental exploration.

To address these limitations, we plan to establish a new dataset that captures images of the same subjects under both controlled and uncontrolled lighting conditions. This paired dataset, combined with advanced deep learning techniques, will enable us to better train and validate color correction methods for tongue image analysis under varying lighting conditions.

\subsection{Expanding Dataset with Attributed Bounding Boxes}
Previous studies have developed tongue image datasets with annotated bounding boxes for specific attributes, such as those presented in~\citesup{chang2024tongue, jiang2022deep_sup}. However, due to ethical considerations and privacy regulations, these datasets are not publicly available, limiting their broader application in research and development.

\begin{figure}[tb]
  \centering
  \includegraphics[width=0.9\linewidth]{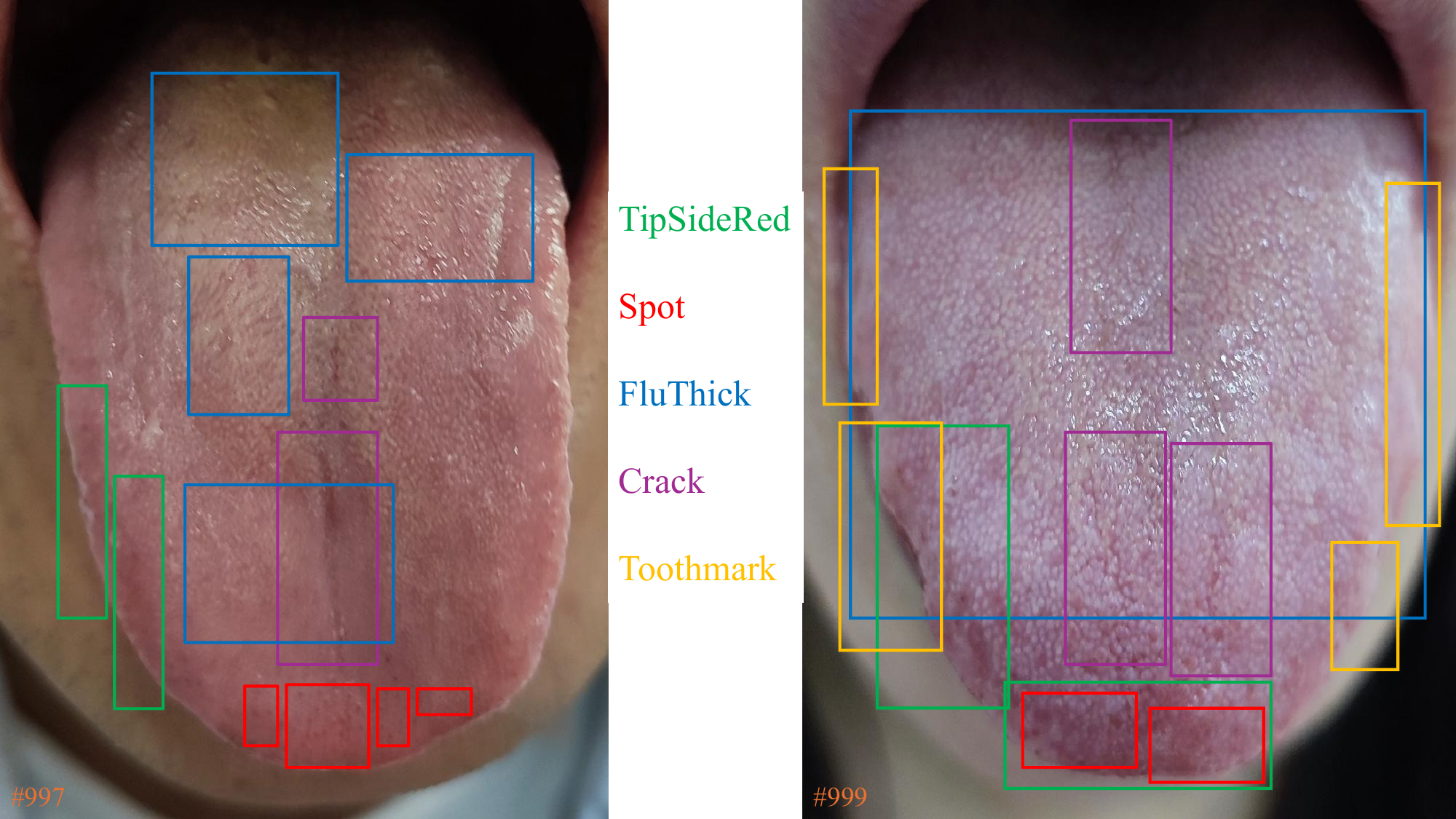}
  \caption{Examples of tongue images with bounding box annotations for different tongue attributes. Different colored boxes represent various attributes: TipSideRed (blue), Spot (red), FluThick (green), Crack (black), and Toothmark (yellow).}
  \label{fig:7}
\end{figure}

To address this gap, our future work includes plans to enhance our existing dataset by incorporating bounding box annotations that specify the precise locations of different attributes on the tongue surface. We have already generated some preliminary samples with such annotations in Fig.\ref{fig:7}. We believe that these spatially-explicit annotations will significantly improve our attribute recognition capabilities, as the current performance metrics shown in Tables~\ref{tab:3} and~\ref{tab:4} indicate that our method's accuracy falls short of the standards required for practical clinical applications. These bounding box annotations would provide more fine-grained supervision during training, potentially leading to more precise and clinically relevant attribute recognition.

\subsection{Video-Based Tongue Attributes Dataset}
While the comparisons presented in Tables~\ref{tab:3} and~\ref{tab:4} offer a fair evaluation of our method, they inadvertently place human evaluators at a significant disadvantage. In real remote clinical settings, practitioners observe dynamic tongue movements rather than being limited to static images. The challenge of attribute recognition from a single static image is substantial, as evidenced by the near-random performance (approximately 50\%) of laypersons and the relatively modest accuracy rates even among practitioners with 10 years of diagnostic experience.

This observation underscores the necessity of developing a video-based tongue attribute dataset. Such a dataset would better simulate the actual remote diagnosis process, where practitioners can observe dynamic tongue movements and select optimal frames that are least affected by environmental factors for their final assessment. This approach would align more closely with real-world remote clinical practice, where temporal information plays a crucial role in accurate diagnosis. We plan to establish such a comprehensive video-based tongue dataset, enabling our networks to better emulate the diagnostic process of experienced practitioners and potentially achieve more reliable attribute recognition in telemedicine applications.

\subsection{Multi-Attribute Learning and Cross-Feature Enhancement}
Beyond the eight fundamental tongue attributes analyzed in this study, our dataset encompasses additional annotations including demographic information (age, gender) and various TCM diagnostic features. While these auxiliary attributes were not utilized in our current experiments, they represent valuable information that could be leveraged in future research like ~\citesup{ho7visceral}. One promising direction would be to explore how these supplementary attributes could be incorporated into the training process to enhance recognition accuracy through mutual feature reinforcement. For instance, certain tongue attributes may exhibit correlations with age groups or gender, or demonstrate established relationships with specific TCM diagnostic patterns. By incorporating these interconnected features into our learning framework, we could potentially develop a more holistic and robust attribute recognition system.

\bibliographystylesup{aaai25}
\bibliographysup{sup}

\end{document}